\documentstyle[prc,aps,multicol]{revtex}

\begin{document}

\draft

\title{Nuclear and Coulomb Interaction in the $^8$B $\rightarrow$ $^7$Be + p  
Breakup  Reaction at  sub-Coulomb Energies}

\author{ V. Guimar\~aes , J. J. Kolata, D. Peterson, P. Santi, 
R. H. White-Stevens and  S. M. Vincent}
\address{Physics Department, University of Notre Dame, Notre Dame,
Indiana 46556-5670 USA}

\author{F. D. Becchetti, M. Y. Lee, T. W. O'Donnell, D. A. Roberts and 
J. A. Zimmerman.}
\address{ Physics Department, University of Michigan, Ann Arbor,
Michigan 48109-1120 USA}

\date{\today}

\maketitle

\begin{abstract}

The angular distribution for the breakup of  
$^8$B~$\rightarrow$~$^7$Be~+~p on a $^{58}$Ni target has been measured
at an incident energy of 25.75 MeV.  The data are inconsistent with 
first-order theories but are remarkably well described by calculations
including higher-order effects. The comparison with  theory illustrates
the importance of  the exotic  proton halo structure of $^8$B in accounting
for the observed breakup angular distribution. 
\end{abstract}


\begin{multicols}{2}

Coulomb dissociation reactions have been used in recent years as a 
means to obtain information on  capture reactions of astrophysical 
interest. 
 An example is the experiment of  Motobayashi, et al. \cite{mot94} who 
studied the breakup of $^8$B~$\rightarrow$~~$^7$Be~+~p on a Pb target
 and related their result to  radiative proton capture at solar energies. 
This reaction corresponds to the projectile
breaking up into a  core and a valence nucleon
due to interactions with virtual photons in the strong Coulomb field of
a high-Z nucleus.  
Although this mechanism  
is in principle the time reversal of a capture reaction, 
 $E2$ photons can contribute significantly to  Coulomb dissociation 
while radiative  capture at solar energies 
 proceeds almost exclusively by $E1$ transitions.
Thus, in extracting information  on  
astrophysical proton capture reactions from the measured dissociation
cross section it is  crucial to determine the relative contribution
of  photons having different multipolarity.

The relative importance of $E1$ and $E2$ contributions to the Coulomb 
dissociation of $^8$B has been investigated both experimentally 
\cite{sch96,kik97,dav98,iwa99} and theoretically \cite{lan94,gai95,esb96}.  
The earliest experiments \cite{sch96,kik97} suggested that the $E2$ strength 
was much smaller than all published theoretical estimates.  Davids, et al. \cite{dav98} measured the asymmetry in the 
longitudinal momentum distribution of $^7$Be fragments from the dissociation
of $^8$B on Pb at 44 and 81 MeV per nucleon.  Their high-energy data were
quite ambiguous, but the 44 MeV/nucleon results gave a clear 
signal corresponding to an $E2$ strength that was 70$\%$ of that predicted
by the model of Esbensen and Bertsch \cite{esb96}.  This model prediction
itself is a factor of two smaller than that of Kim, et al. \cite{kim87}. 
Nevertheless the extracted $E2$ strength, though considerably quenched, is 
still larger than implied in Refs. \cite{sch96} and \cite{kik97}.  Most 
recently, Iwasa, et al. \cite{iwa99} report a limit on the $E2$ strength
that is at least an order of magnitude smaller than that of Davids, et al.. 

It was noted in Ref. \cite{dav98} that the description of the data by the
model of Esbensen and Bertsch is not precise, and that the best-fit values for the $E1$ and $E2$ strengths differ by 20-30$\%$ from the model predictions.
The $E2/E1$ interference term is, of course, model dependent.  
The earlier experiment of von Schwarzenberg, et al. 
\cite{sch96} had attempted to avoid model dependence by measuring the breakup
at sub-Coulomb energies for a low-Z target  ($^{58}$Ni) for which multiple
Coulomb excitation was expected to be minimal.  At these energies, the
E2 component is enhanced relative to E1.  The very small cross section
reported in that work, 
which was less than that predicted by any reasonable structure model for $^8$B
\cite{nun98},  has generated considerable interest.  Nunes and Thompson
\cite{nun98} and Dasso, Lenzi, and Vitturi \cite{das98} independently
suggested that the explanation for this result might be strong destructive
nuclear-Coulomb interference effects, despite the 
fact that at the angle where the measurement was made the classical distance 
of closest approach  is nearly 20 fm, i.e., far outside the range of the nuclear force 
for a ``normal'' nuclear system.   A strong nuclear-dominated
peak in the differential  cross section  at a center-of-mass angle 
of $70^{\rm o}-90^{\rm o}$  (well inside the expected 
$100^{\rm o}-110^{\rm o}$ for the onset of nuclear breakup of a normal nucleus)
was predicted in Refs.  \cite{nun98} and \cite{das98}, although it 
was pointed out that the corresponding calculations are only first-order in 
the nuclear and Coulomb fields and might be modified by multi-step excitations.   
 Furthermore, it was suggested in
Ref. \cite{nun98} that even pure Coulomb excitation would be considerably
modified from that expected in the normal ``point-Coulomb'' approximation
which ignores the extended size of the valence proton wave-function in 
$^8$B (see Ref. \cite{nun98} for a more complete discussion of this approximation).  This leads to a further reduction in the calculated breakup cross section. 
Since both these effects are directly attributable to the 
exotic ``halo" structure of $^8$B, it is important to verify, if possible, 
the implications of these theoretical calculations. 
The present work  was carried out in an attempt to test these predictions 
by obtaining angular distribution for the breakup of 
$^8$B on $^{58}$Ni at the same incident energy as the previous 
experiment \cite{sch96}, but over as much of  the critical angular range as possible.

	The experiment was carried out at the Nuclear Structure
Laboratory of the University of Notre Dame.
To produce  the low-energy secondary radioactive $^8$B beam, we used
the {\it TwinSol} radioactive ion beam (RIB) facility \cite{lee99}
and the $^6$Li($^3$He,n)$^8$B
 direct transfer reaction. A gas target containing
1 atm of $^3$He was bombarded by a high-intensity (up to 300
particle-nA), nanosecond-bunched
primary $^6$Li beam
at an energy of 36 MeV.  The entrance and exit windows of the gas cell
consisted of 2.0 $\mu$m Havar foils.  The secondary beam was selected
and transported through the solenoids and then
focused onto a  924 $\mu$g/cm$^2$ thick, isotopically-enriched $^{58}$Ni
secondary target. The laboratory energy of the $^8$B beam
at the center of this target was 25.75 MeV, with
an overall resolution of 0.75 MeV full width at half maximum (FWHM)
and an intensity of 2.5 $\times$ 10$^4$ particles per second.
The spread in energy was mainly due to a combination of the
kinematic shift in the production reaction and the energy-loss
straggling in the gas cell windows and the  $^{58}$Ni  target.
The beam had a maximum angular divergence of  $\pm$~$4^{\rm o}$
and a spot size of approximately 4.0 mm FWHM. Although the
count rate in the detectors was modest (typically less than
2$\times$ 10$^3$ s$^{-1}$), the expected breakup yield is low
so pulse-pileup tagging with a resolving time of 50 ns was 
used to eliminate pileup events.

The $^8$B breakup events,  and also elastically-scattered particles,
were detected with two telescopes consisting of 25 and 30 $\mu$m Si $\Delta$E
detectors, backed  by  thick Si E detectors. These were
placed on either side of the beam at $\Theta_{\rm LAB} = $
$20^{\rm o}, 30^{\rm o}, 40^{\rm o}, 45^{\rm o}, 
 50^{\rm o}$ and $ 60^{\rm o}$.
Each telescope had a circular collimator that subtended a solid
angle of 41~msr, corresponding to  a overall effective angular resolution 
of  10.9$^{\rm o}$ (FWHM), computed by folding in the acceptance 
of the collimator with the spot size and angular divergence of the beam. 

Unambiguous separation of the $^7$Be fragments resulting from $^8$B breakup 
from $^7$Be contamination in the direct beam elastically scattered by the 
 $^{58}$Ni  target was crucial to the success of this experiment. 
 Although contaminants 
were present in the beam, they could be identified using time-of-flight (TOF)
techniques.  The TOF of the particles was obtained from the time difference 
between the occurrence of an E signal in a telescope and the RF timing pulse 
from the beam buncher.  The time resolution of better than 3 ns (FWHM) was 
 adequate to separate $^7$Be from $^8$B, as illustrated in Fig. 1.  
At  sub-Coulomb energies, it is not easy to 
carry out a coincidence measurement between the $^7$Be 
fragment and the proton, as was done at higher energies \cite{mot94,kik97}, 
due to  the much reduced kinematic focusing of the protons. Such 
an experiment requires 4$\pi$ geometry
and the ability to detect very low-energy protons to avoid biasing the 
correlation.  Thus, we determined only the  integrated $^7$Be yield 
from the dissociation  reaction $^8$B $\rightarrow$ $^7$Be + p. 
 Although the contaminants in  the secondary $^8$B beam are well
separated in the $\Delta$E vs. E$_{\rm TOTAL}$ spectrum, as illustrated in
Fig. 1(a),  it would not have been possible to separate the $^7$Be 
products coming from  breakup events from the scattered contaminant $^7$Be 
beam using only this information.  However, by also considering  the
TOF  information,  particles of different origins could be
completely separated as shown in Fig. 1(b), since the $^7$Be from 
$^8$B breakup has the same TOF  as the $^8$B beam.

The experimental angular distribution deduced for the
dissociation of  $^8$B into $^7$Be + p on a $^{58}$Ni target is presented 
in Figs.  2 and 3, as a function of the center-of-mass angle of the detected $^7$Be 
(we used the $^8$B elastic-scattering Jacobian
to  transform the laboratory angles to the center-of-mass frame).
The differential cross sections were obtained
by integrating  $^7$Be breakup events over the solid angle
subtended by the two telescopes.  The number of $^8$B ions per 
integrated charge of the primary  beam was determined in a separate run.
The normalization was obtained 
using the information on solid angle, target thickness and 
the measured integrated charge of the primary beam 
for each run,  and verified by  a measurement of  $^8$B 
elastic scattering, which is expected to be purely Rutherford at 
forward angles.  The systematic error in the absolute normalization is estimated 
to be approximately 10$\%$, mainly due to the uncertainty in the intensity
of the secondary beam. The  $^8$B beam had a $1^{\rm o}$ angular 
offset from the center axis set for the telescopes. This shift, 
evaluated by analyzing the observed asymmetry in the elastic scattering 
of $^8$B, had a strong effect on the differential cross section at forward 
angles. Thus, at the most-forward angle setting of the telescopes 
we display the differential cross sections 
obtained at $\Theta_{\rm LAB}$ =~19$^{\rm o}$~and~21$^{\rm o}$ separately.  
At backward angles, where the cross section does not 
change so rapidly  as a function  of angle,  we have taken 
the average of the yield  measured in the two telescopes.

It is obvious from inspection of the experimental angular distribution 
(see Fig. 3) that our data are completely inconsistent with the large 
amplitude peak in the vicinity of 70$^{\rm o}$ - 90$^{\rm o}$ which was 
a prominent feature of both first-order theories 
\cite{nun98,das98}.  Very recently, however, two calculations 
\cite{esb99,nun99} that incorporate higher-order effects have 
been published, and they display a much different large-angle behavior.  Esbensen and Bertsch \cite{esb99} performed a dynamical calculation that followed the 
time evolution of the valence proton wave function to all orders in the 
Coulomb and nuclear fields of the target.  
Their results are compared with our data in Fig. 2.  The dotted 
curve corresponds to pure Coulomb breakup while the dashed curve, 
which includes nuclear effects, can be directly 
compared with the calculations presented in Refs. \cite{nun98} and 
\cite{das98}.  It can be seen that the higher-order couplings have 
completely eliminated the large-angle peak predicted by these first-order 
theories.  Nevertheless, it is also clear that 
Coulomb-nuclear interference at very large distances, due to the 
extended nature of the ``proton halo'' in $^8$B, still plays an important role 
in accounting for the experimental data.  

Two other curves also appear in Fig. 2.  The thin solid line 
illustrates pure Coulomb excitation under the usual ``point-like'' assumption. 
The dotted curve, which is much closer to the experimental data, is the
correct pure-Coulomb-excitation calculation which takes account of the
extended size of the valence proton orbital of the projectile.  
This result emphasizes the importance of incorporating the unusual structure 
of $^8$B in all aspects of the reaction dynamics as first discussed in 
Ref. \cite{nun98}. (Note that the ``point-like'' approximation is still
valid for neutron-halo nuclei since the relevant distance in this
case is that between the core and the center-of-mass of the halo nucleus, which is still small). The thick solid curve includes, in addition to breakup, 
the effect of nucleon 
transfer from the projectile to the target.  This is the calculation that is 
most appropriate for comparison with our data since we do not distinguish 
transfer from breakup.  The large-angle peak is partially restored 
(but transfer was not included in the calculations presented in Refs. \cite{nun98} and \cite{das98} so the computed transfer yield should be
added to the angular distributions presented there).  The present data suggest that proton transfer may have been somewhat overestimated in Ref. \cite{esb99}.  
Nevertheless, the overall agreement between theory 
and experiment is remarkable, especially considering  that there has been
no renormalization of the predicted absolute cross section.

Nunes and Thompson \cite{nun99} have also included higher-order effects, 
using the coupled discretized continuum channels (CDCC) method combined with 
the structure model of Esbensen and Bertsch \cite{esb96}.  The advantage of this 
approach was that they were able to explicitly show that the vanishing of the 
large-angle peak results directly from the coupling among continuum states.  
Nunes \cite{nun99b} has added proton transfer to this calculation  and the 
result appears as the thick solid line in Fig. 3.  She has also repeated the 
calculation using the structure model of Kim, et al. \cite{kim87} which, 
as mentioned  above, has both a larger $E1$ and $E2$ component.  
The result is shows as the 
thin solid curve in Fig. 3.  In general, the data favor the CDCC calculation 
using the wave function of Ref. \cite{esb96}, but the differences are small. 

In conclusion, the angular distribution of the breakup of $^8$B 
into $^7$Be + p  on a $^{58}$Ni target was measured over a wide range 
of angles at a laboratory energy of 25.75 MeV.  Time-of-flight information
allowed us to unambiguously separate the $^7$Be fragments coming from the 
breakup process, considerably improving on a previous measurement 
\cite{sch96}. 
The  data are completely inconsistent with first-order reaction theories 
\cite{nun98,das98}  which predict a large amplitude nuclear dominated  
peak in the cross section at a center-of-mass angle of $70^{\rm o}-90^{\rm o}$.  
However, recent calculations \cite{esb99,nun99,nun99b} 
incorporating higher-order effects are in excellent agreement with experiment.
In these calculations, the spurious peak is eliminated by continuum-continuum 
couplings.   Coulomb-nuclear interference at very large 
distances, and the need to account for the extended size of the valence proton 
wave function in computing Coulomb breakup, are important features of both 
calculations.  Thus, the present data may well be the best evidence yet of an 
exotic ``proton halo'' structure for $^8$B.  This has been a matter of some 
controversy, since reaction cross section measurements at relativistic 
energies by Tanihata, et al. \cite{tan85} displayed little or no 
enhancement, while similar measurements at intermediate energies by 
Warner, et al.  \cite{war95} and Negoita, et al. \cite{neg96} showed a rather 
substantial enhancement.  (Enhancements in the reaction cross sections 
were the first signature of the neutron halo).  The present data illustrate 
that finite-size effects and nuclear-Coulomb interference at very large 
distances, well outside the ``normal'' range of the nuclear force, are crucial features for the understanding of $^8$B reactions at near-barrier and sub-barrier energies.

The original goal of the experiment described in Ref. \cite{sch96} was to 
obtain a model-independent measure of the $E2$ component in $^8$B breakup and 
the astrophysical S-factor S$_{17}$ for proton capture 
on $^7$Be at solar energies.  In light of the discussion above, 
it appears that this will be very difficult. 
 Even at the farthest forward angles measured in this experiment, 
 corresponding to a distance of closest approach greater than 30 fm, 
 substantial Coulomb-nuclear (and multiple-Coulomb-excitation) 
interference effects
are important.  While our data are consistent with the results of 
 Davids, et al. \cite{dav98}, in the sense that the same structure model 
 provides good predictions for both data sets, this 
conclusion is model dependent.  On the other hand, 
the results from Ref. \cite{dav98} are also model
dependent and the applicability of first-order 
perturbation theory and the ``point-Coulomb'' approximation, 
used there and in the 
analysis of breakup data at intermediate energies \cite{mot94,kik97}, 
should be re-investigated.
 
There does appear to be some sensitivity to the various structure models  
in our data. The wave function of Kim, et al. \cite{kim87}, which has the larger
S$_{17}$ and $E2$ components, does not fit the data as well as other
models, but the differences are  too small to allow us to make any definitive 
statements about either quantity at this time.

Finally, the interactions of exotic, weakly-bound nuclei at near- and 
sub-barrier energies will increasingly be investigated as the next generation of 
radioactive ion beam facilities using the ISOL technique become available.  
It is comforting that there exist at least two successful theoretical 
approaches to the difficult problem of understanding low-energy reaction 
dynamics of weakly bound nuclei. We have shown that the information obtained 
from these reactions is complementary to that obtained 
from studies at intermediate and relativistic energies.  

One of us  (V.G.)  was financially 
supported by FAPESP (Funda\c c\~ao de 
Amparo a Pesquisa do Estado de S\~ao Paulo - Brazil) while on leave 
from the UNIP (Universidade Paulista). 
This work was supported by the National Science Foundation under 
Grants No. PHY94-01761, PHY95-12199, PHY97-22604, 
PHY98-04869 and PHY99-01133.

\end{multicols}
\vspace{5cm}

\begin{figure}
\caption{{\bf (a)} The $\Delta$E vs. E$_{\rm TOTAL}$ spectrum taken at 
$\Theta_{\rm LAB}~=~45^{\rm o}$. The $^7$Be and $^8$B gates are shown.
{\bf (b)} TOF-$\Delta$E  spectrum illustrating the separation between
the $^7$Be breakup events and elastically-scattered $^7$Be in the 
direct beam. This spectrum corresponds to events in the gates shown in 
Fig. 1(a). The breakup events are emphasized with larger dots. The 
energy calibration and time calibration are 20 keV/channel and 
0.50 ns/channel, respectively.}
\end{figure}

\begin{figure}
\caption{Experimental angular distribution for $^8$B breakup as measured 
  in this work, compared with the calculations presented in Ref. \protect\cite{esb99}. 
  The various curves are discussed in the text.}
\end{figure}

\begin{figure}
\caption{The experimental data compared with the calculation of Nunes, et al.
  (Ref. \protect\cite{nun99,nun99b}).  
  The various curves are discussed in the text, 
  except for the two dot-dashed curves. These are the separate contributions 
  of  transfer reactions as  calculated in the structure model of  
  Ref. \protect\cite{esb96} and Ref. \protect\cite{kim87}.  The dashed
  curve with peaks at 15$^{\rm o}$ and 85$^{\rm o}$ is the first-order
  calculation from Ref. \protect\cite{nun98}.}
\end{figure}

\end{document}